\documentclass[preprint,aps,prd]{revtex4}
\usepackage{epsfig}

\RequirePackage{xspace}
\usepackage{epsfig}
\usepackage{graphicx}
\usepackage{color}

\newcommand{\dv}{\partial\hspace{-7pt}\slash}

\newcommand{\be}{\begin{equation}}
\newcommand{\ee}{\end{equation}}
\newcommand{\bea}{\begin{eqnarray}}
\newcommand{\eea}{\end{eqnarray}}

\begin{document}
\title{Lepton FCNC in Type III Seesaw Model}
\author{Xiao-Gang He$^{1,2}$}\email{hexg@phys.ntu.edu.tw}
\author{Sechul Oh$^{1}$}\email{scoh@phys.ntu.edu.tw}
\affiliation{ $^1$Department of Physics and Center for Theoretical
Sciences, National Taiwan University,
Taipei, Taiwan\\
$^2$Center for High Energy Physics, Peking University, Beijing 100871, China}


\begin{abstract}
In Type III seesaw model, there are tree level flavor changing neutral currents (FCNC) in the
lepton sector, due to mixing of charged particles in the leptonic triplet introduced to realize
seesaw mechanism, with the usual charged leptons. In this work we study these FCNC effects in
a systematic way using available experimental data. Several FCNC processes have been studied
before. The new processes considered in this work include:
lepton flavor violating processes
$\tau \to P l$, $\tau \to V l$, $V \to l \bar l'$, $P \to l \bar l'$, $M\to M' l \bar l'$ and
muonium-antimuonium oscillation. Results obtained are compared with previous results from
$l_i \to l_j l_k \bar l_l$, $l_i \to l_j \gamma$, $Z \to l \bar l'$ and $\mu - e$ conversion.
Our results show that the most stringent constraint on the $e$-to-$\tau$ FCNC effect comes from
$\tau \to \pi^0 e$ decay. $\tau \to \rho^0 \mu$ and $\tau \to \pi^0 \mu$ give very stringent
constraints on the $\mu$-to-$\tau$ FCNC effect, comparable with that obtained from
$\tau \to \mu \bar \mu \mu$ studied previously. The constraint on the $e$-to-$\mu$ FCNC effect
from processes considered in this work is much weaker than that obtained from processes studies
previously, in particular that from $\mu - e$ conversion in atomic nuclei.
We find that in the canonical seesaw models the FCNC parameters, due to tiny neutrino masses,
are all predicted to be much smaller than the constraints obtained here, making such models
irrelevant.  However, we also find that in certain special circumstances the tiny neutrino masses
do not directly constrain the FCNC parameters. In these situations, the constraints from the FCNC
studies can still play important roles.
\end{abstract}


\maketitle


\noindent
{\bf Introduction}

Neutrino oscillation experiments involving neutrinos and antineutrinos coming from astrophysical
and terrestrial sources have found compelling evidence that neutrinos have finite but small masses.
To accommodate this observation, the minimal standard model (SM) must be extended. Some sensible
ways  to do this include:
(a) Type I seesaw with three heavy right-handed (RH) Majorana neutrinos \cite{type1_seesaw},
(b) the use of an electroweak Higgs triplet to directly provide the left-handed (LH)
neutrinos with small Majorana masses (Type II seesaw \cite{type2_seesaw}),
(c) introducing  fermion triplets with zero hypercharge (Type III seesaw \cite{type3_seesaw}),
(d) the generation of three Dirac neutrinos through an exact parallel of
the SM method of giving mass to charged fermions,  and (e) the radiative generation of neutrino
masses as per the Zee or Babu models \cite{zeebabu}.
But in the absence of more experimental data, it is impossible to tell which, if any, of these
is actually correct.  Different models should be studied using available data or future ones.
In this work, we carry out a systematic study of constraints on possible new flavor changing neutral
currents (FCNC) in Type III seesaw model.

The fermion triplet $\Sigma$ in Type III seesaw model transforms
under the SM gauge group $SU(3)_C\times SU(2)_L\times U(1)_Y$ as (1,3,0).
We will assume that there are three copies of such fermion triplets.
The model has many interesting features~\cite{model}, including the possibility of having low seesaw scale
of order a TeV to realize leptogenesis~\cite{leptogenesisIII} and detectable effects at
LHC~\cite{production,he-chen} due to the fact that the heavy triplet leptons have gauge interactions
being non-trivial under the $SU(2)_L$ gauge group,
and the possibility of having new tree level FCNC interactions in the lepton
sector~\cite{fcnc, fcnc2, fcnc3}. Some of the FCNC effects have been studied, such as
$l_i \to l_j \bar l_k l_l$, $l_i \to l_j \gamma$, $Z \to l_i \bar l_j$ and $\mu - e$ conversion processes.
Several other FCNC processes studied experimentally have not been studied in the context of
Type III seesaw model.
We will study constraints on FCNC related to charged leptons in a systematic way using available
experimental bounds listed in Ref.~\cite{pdg} by the particle data group.

Before studying constraints, let us describe the model in more detail to identify new tree level FCNC
in the lepton sector.  The component fields of the righthanded triplet $\Sigma$ are,
\begin{eqnarray}
\Sigma&=&
\left(
\begin{array}{ cc}
   N^0/\sqrt{2}  &   E^+ \\
     E^- &  -N^0/\sqrt{2}
\end{array}
\right), \quad
\Sigma^c=
\left(
\begin{array}{ cc}
   N^{0c}/\sqrt{2}  &   E^{-c} \\
     E^{+c} &  -N^{0c}/\sqrt{2}
\end{array}
\right)\;,
\end{eqnarray}
and the renormalizable Lagrangian involving $\Sigma$ is given by
\begin{equation}
\label{Lfermtriptwobytwo}
{\cal L}=Tr [ \overline{\Sigma} i \slash \hspace{-2.5mm} D  \Sigma ]
-\frac{1}{2} Tr [\overline{\Sigma}  M_\Sigma \Sigma^c
                +\overline{\Sigma^c} M_\Sigma^* \Sigma]
- \tilde{H}^\dagger \overline{\Sigma} \sqrt{2}Y_\Sigma L_L
-  \overline{L_L}\sqrt{2} {Y_\Sigma}^\dagger  \Sigma \tilde{H}\, ,\nonumber
\end{equation}
where $L_L  = (\nu_L, l^-_L)^T$ is the lepton doublet. $H\equiv (\phi^+, \phi^0)^T \equiv
(\phi^+, (v+h+i \eta)/\sqrt{2})^T$ is the Higgs doublet with $v$ being the vacuum expectation
value, and $\tilde H = i \tau_{2} H^*$.

Defining $E\equiv E_R^{+ c} + E_R^-$ and removing the would-be Goldstone bosons $\eta$ and
$\phi^\pm$, one obtains the Lagrangian
\begin{eqnarray}
\label{Lfull-ft-2}
{\cal L}&=& \overline{E} i \dv E
+ \overline{N_R^0} i \dv  N^0_R
-  \overline{E}M_\Sigma E -
        \left( \overline{N^{0}_R} \frac{{M_\Sigma}}{2}  N_R^{0c} \,+  \,\text{h.c.}\right)
\nonumber \\
&+&g \left(W_\mu^+ \overline{N_R^0} \gamma_\mu  P_R E
 +  W_\mu^+ \overline{N_R^{0c}} \gamma_\mu  P_L E   \,+  \,\text{h.c.}
 \right) - g\, W_\mu^3 \overline{E} \gamma_\mu  E
 \nonumber\\
&-&  \left( {1\over \sqrt{2}}(v+h) \overline{N_R^0} Y_\Sigma \nu_{L}+ (v + h) \overline{E} Y_\Sigma
l_{L} \,+  \,\text{h.c.}\right)\,.
\end{eqnarray}

One can easily identify the terms related to neutrino masses from the above. The mass matrix is the
seesaw form
\bea
{\cal L}= -(\overline{\nu_L^c}\,\, \overline{N^{0}} )\left(
\begin{array}{ cc}
  0  &   {Y_\Sigma}^T v/2\sqrt{2} \\
   {Y_\Sigma} v/2\sqrt{2} &  {M_\Sigma}/2
\end{array}
\right)
\left(
\begin{array}{ c}
   \nu_L \\
   N^{0c}
\end{array}
\right) + {\rm h.c.}\;.
\label{neutralfullmassmatrix}
\eea

The charged partners in the triplets mix with the SM charged leptons resulting in
a mass matrix of the following form
\begin{equation}
{\cal L} = -(\overline{l_R}\,\, \overline{E_R} )
\,\,
\left(
\begin{array}{ cc}
   m_l  &   0 \\
      {Y_\Sigma} v &  {M_\Sigma}
\end{array}
\right) \,\,
\left(
\begin{array}{ c}
   l_L \\
  E_L
\end{array}
\right) + {\rm h.c.}\;.
\end{equation}

One can diagonalize the fermion mass matrices and find the transformation matrices between fields
in weak interaction basis and in mass eigenstate basis defined as
\begin{equation}
\left(
\begin{array}{ c}
   l_{L,R} \\
  E_{L,R}
\end{array}
\right) = U_{L,R}
\left(
\begin{array}{ c}
   l'_{L,R} \\
   E'_{L,R}
\end{array}
\right)\,,\;\;\;\; \left(
\begin{array}{ c}
   \nu_{L} \\
   N^{0c}
\end{array}
\right) = U
\left(
\begin{array}{ c}
   \nu'_{L} \\
   N'^{0c}
\end{array}
\right)\;,\nonumber
\end{equation}
where the primed fields indicate mass eigenstates. $U_{L,R}$ are $(3+3)$-by-$(3+3)$ matrices
if $3$ triplets are present, and can be written as
\begin{equation}
U_{L}\equiv
\left(
\begin{array}{ cc}
   U_{Lll} &   U_{L lE} \\
     U_{LE l} & U_{L EE}
\end{array}
\right) \,,\,
U_{R}\equiv
\left(
\begin{array}{ cc}
   U_{Rll} &   U_{R lE} \\
     U_{RE l} & U_{R EE}
\end{array}
\right) \,,\,
U_{}\equiv
\left(
\begin{array}{ cc}
   U_{\nu \nu } &   U_{\nu N} \\
     U_{N \nu} & U_{NN}
\end{array}
\right)\,.\\
\label{mixdef}
\end{equation}
To order $v^2M^{-2}_\Sigma$, one has~\cite{fcnc}
\begin{eqnarray}
\label{epsilon}
&& U_{Lll} = 1- \epsilon\;,\;\;U_{LlE} = Y^\dagger_\Sigma M^{-1}_\Sigma v\;,\;\;\;\;\;\;\;
U_{LE l} = - M^{-1}_\Sigma Y_\Sigma v\;,\;\;\;\;\;\;U_{LEE} = 1-\epsilon'\;,\nonumber\\
&& U_{Rll} = 1\;,\;\;\;\;\;\;\;\;U_{RlE} = m_l Y^\dagger_\Sigma M^{-2}_\Sigma v\;,\;\;\;
U_{REl} = - M^{-2}_\Sigma Y_\Sigma m_l v\;,\;\;U_{LEE} = 1\;,\nonumber\\
&& U_{\nu\nu} = (1- \epsilon/2)U_{PMNS}\;, ~~~~~~~~~~~~~
U_{\nu N} = Y^\dagger_\Sigma M^{-1}_\Sigma v/\sqrt{2}\;,\;\; \nonumber\\
&& U_{N \nu} = - M^{-1}_\Sigma Y_\Sigma U_{\nu\nu} v/\sqrt{2}\;, ~~~~~~~~~
U_{NN} = 1-\epsilon'/2\;,
\end{eqnarray}
where
\begin{eqnarray}
\label{epsilon1}
\epsilon = Y^\dagger_\Sigma M^{-2}_\Sigma Y_\Sigma v^2/2 = U_{\nu N} U_{\nu N}^{\dagger} ~~~~~~
\epsilon' = M^{-1}_\Sigma Y_\Sigma Y^\dagger_\Sigma M^{-1}_\Sigma v^2/2
 = U_{\nu N}^{\dagger} U_{\nu N} \;.
\end{eqnarray}
Here $U_{PMNS}$ denotes the lowest order Pontecorvo-Maki-Nakagawa-Sakata (PMNS) mixing matrix which
is unitary.  We have kept higher order corrections to the $U_{\nu \nu}$ matrix.

Using the above, one obtains the couplings of Z and physical Higgs $h$ to the usual charged leptons
\bea
\label{NC}
\mathcal{L}_{NC}&=&\frac{g}{cos\theta_W}\overline{l} \gamma^{\mu}
\left(P_L (- \frac{1}{2}+ \sin^2\theta_W-\epsilon) +P_R \sin^2\theta_W \right) l Z_\mu\;,\nonumber\\
%
\label{Heta}
\mathcal{L}_{H}&=&\frac{g}{2M_W}\overline{l}\left(P_L m_l\left(3\epsilon-1\right)
+P_R \left(3\epsilon-1\right)m_l\right)l h\;.
\eea
Here we have dropped the ``prime'' on the fermion mass eigenstates.
$\epsilon$ is a 3-by-3 matrix. Non-zero off diagonal elements in $\epsilon$ are the new sources
of tree level FCNC in charged lepton sector. The $Z$ and Higgs coupling to quarks are the same as in the SM.
We will use available FCNC data in a systematic way to constrain the parameter $\epsilon_{ll'}$.

Several processes, such as
$l_i \to l_j l_k \bar l_l$, $l_i \to l_j \gamma$, $Z \to l \bar l'$ and $\mu - e$ conversion in atomic
nuclei, have been studied and stringent constraints have been obtained for $\epsilon_{ll'}$ which will
be used as standards for constraints obtained from new lepton flavor violating (LFV) processes considered
here, $\tau \to P l$, $\tau \to V l$, $V \to l \bar l'$, $P \to l \bar l$, $M\to M' l \bar l'$
and muonium-antimuonium oscillation.
It turns out that with currently available experimental data, the LFV processes considered in this work
involving $\tau$ leptons provide very stringent constraints on the FCNC parameter $\epsilon_{i \tau}$.
Our results show that the most stringent constraint on $\epsilon_{e\tau}$ comes from $\tau \to \pi^0 e$
decay. $\tau \to \rho^0 \mu$ and $\tau \to \pi^0 \mu$ give very stringent constraints on $\epsilon_{\mu\tau}$,
comparable with that obtained from $\tau \to \mu \bar \mu \mu$ in previous studies. The strongest constraint
on $\epsilon_{e\mu}$ comes from $\mu - e$ conversion in atomic nuclei studied previously.
We now present some details for the new processes mentioned above.
\\


\noindent
{\bf Constraints from $\tau \to P l$ and $\tau \to V l$}

Exchange of $Z$ boson between quarks and leptons can induce $\tau \to P l$ and $\tau \to V l$
at tree level, where a pseudoscalar meson $P = \pi^0, \eta, \eta'$ and a vector meson
$V = \rho^0, \omega, \phi$ and a charged lepton $l = e, \mu$.
The decay amplitudes for $\tau \to M l$ (where $M$ denotes either $V$ or $P$) can be written in the
following form
\begin{eqnarray}
\label{Tau-decay-amp}
{\cal M} &=& 2\sqrt{2} G_F ~\epsilon_{l \tau}
\sum_{q = u, d, s}  \langle  M(p_{_M}) | \bar q \gamma_{\alpha} (I_3 P_L - Q_q \sin^2\theta_W) q | 0 \rangle
\cdot \left[ \bar l (p_{_l}) \gamma^{\alpha} (1 -\gamma_5) \tau (p_{\tau}) \right] \nonumber\\
 &=& 2\sqrt{2} G_F ~\epsilon_{l \tau} \sum_{q = u, d, s}
\langle M(p_{_M}) | \bar q \gamma_{\alpha} (g^q_{_V} + g^q_{_A} \gamma_5 ) q |0 \rangle
\cdot \left[ \bar l(p_{_l}) \gamma^{\alpha} (1 - \gamma_5) \tau(p_{\tau}) \right],
\end{eqnarray}
where $G_F$ is the Fermi constant, $Q_q$ is the electric charge of q-quark in unit of proton charge.
$I_3 = 1/2$ and $-1/2$ for up and down type of quarks, respectively.
The factor $g^q_{_V}= {1 \over 4} -{2 \over 3}\sin^2 \theta_W$ and $g^q_{_A} = - {1 \over 4}$
for up type of quarks, and $g^q_{_V} = -{1 \over 4} +{1 \over 3}\sin^2 \theta_W$ and
$g^q_{_A} = {1 \over 4}$ for down type of quarks.
The $p_{\tau}$, $p_{_l}$ and $p_{_M}$ are the momenta of $\tau$, $l$ and $M$, respectively.

For $\tau^- \to \pi^0 l$, the decay constant $f_{\pi}$ is defined as
\begin{eqnarray}
\langle \pi^0 (p_{\pi})| {\bar u} \gamma_{\alpha} \gamma_5 u | 0 \rangle
= - \langle \pi^0 (p_{\pi})| {\bar d} \gamma_{\alpha} \gamma_5 d | 0 \rangle
= - i {f_{\pi} \over \sqrt{2}} (p_{\pi})_{\alpha}
\end{eqnarray}
and its value is $f_{\pi} = 130.4~{\rm MeV}$.
For $\tau^- \to \eta l$ and $\tau^- \to \eta^{\prime} l$, due to the $\eta - \eta^{\prime}$
mixing, the decay constants $f^u_{\eta^{(\prime)}}$ and $f^s_{\eta^{(\prime)}}$ are defined as
\begin{eqnarray}
\langle \eta^{(\prime)} (p_{\eta^{(\prime)}})| {\bar u} \gamma_{\alpha} \gamma_5 u | 0 \rangle
&=& \langle \eta^{(\prime)} (p_{\eta^{(\prime)}})| {\bar d} \gamma_{\alpha} \gamma_5 d | 0 \rangle
= - i f^u_{\eta^{(\prime)}} (p_{\eta^{(\prime)}})_{\alpha} ,  \nonumber \\
\langle \eta^{(\prime)} (p_{\eta^{(\prime)}})| {\bar s} \gamma_{\alpha} \gamma_5 s | 0 \rangle
&=& - i f^s_{\eta^{(\prime)}} (p_{\eta^{(\prime)}})_{\alpha} ,
\end{eqnarray}
where
\begin{eqnarray}
f^u_{\eta} &=& {f_8 \over \sqrt{6}} \cos \theta_8 - {f_0 \over \sqrt{3}} \sin \theta_0 , ~~~~
f^s_{\eta} = -2 {f_8 \over \sqrt{6}} \cos \theta_8 - {f_0 \over \sqrt{3}} \sin \theta_0 , \nonumber \\
f^u_{\eta^{\prime}} &=& {f_8 \over \sqrt{6}} \sin \theta_8 + {f_0 \over \sqrt{3}} \cos \theta_0 , ~~~~
f^s_{\eta^{\prime}} = -2 {f_8 \over \sqrt{6}} \sin \theta_8 + {f_0 \over \sqrt{3}} \cos \theta_0 ,
\end{eqnarray}
with $f_8 = 168 ~{\rm MeV}$, $f_0 = 157 ~{\rm MeV}$, $\theta_8 = -22.2^{\circ}$, and
$\theta_0 = -9.1^{\circ}$~\cite{Feldmann}.

For $\tau^- \to V l$ decays, the decay constants $f_{\rho}$, $f_{\omega}$ and $f_{\phi}$ are
defined by
\begin{eqnarray}
&& \langle \rho^0 (p_{\rho})| {\bar u} \gamma_{\alpha} u | 0 \rangle
= -\langle \rho^0 (p_{\rho})| {\bar d} \gamma_{\alpha} d | 0 \rangle
= {f_{\rho} \over \sqrt{2}} m_{\rho} (\epsilon_{\rho})_{\alpha},   \nonumber \\
&& \langle \omega (p_{\omega})| {\bar u} \gamma_{\alpha} u | 0 \rangle
= \langle \omega (p_{\omega})| {\bar d} \gamma_{\alpha} d | 0 \rangle
= {f_{\omega} \over \sqrt{2}} m_{\omega} (\epsilon_{\omega})_{\alpha},   \nonumber\\
&& \langle \phi (p_{\phi})| {\bar s} \gamma_{\alpha} s | 0 \rangle
= f_{\phi} m_{\phi} (\epsilon_{\phi})_{\alpha},
\end{eqnarray}
where $(\epsilon_V)_{\alpha}$ is the polarization vector of $V$.
We use $f_{\rho} = 205~{\rm MeV}$, $f_{\omega} = 195~{\rm MeV}$ and
$f_{\phi} = 231~{\rm MeV}$~\cite{Ball_decay_const_V}.

Exchange of Higgs boson can also induce $q\bar q$ coupling to $l\bar l'$. However,
Higgs-mediated diagrams do not contribute to $\tau \to P l$ and $\tau \to V l$ because
the bi-quark operator in this case is of the form $\bar q q$ which induces a vanishing matrix
element for $<P ~({\rm or}~V)|\bar q q|0>$.

The decay rate for $\tau^- \to P l$ $(P = \pi^0, \eta, \eta^{\prime}$, and $l = e,~ \mu)$, averaged
over the spin of $\tau$ and summed over the spin of $l$, is given by
\begin{eqnarray}
\Gamma = a_P {G_F^2 f_{P}^2 \over 2 \pi m_{\tau}^2} ~| \epsilon_{l
\tau} |^2 \left| {\vec p}_{_{l}} \right| \left[ m_{\tau}^4 + m_l^4
-2 m_l^2 m_\tau^2 -(m_l^2+ m_{\tau}^2 )m_{P}^2) \right],
\end{eqnarray}
where $|\vec p_{_l}| = \sqrt{(m_{\tau}^2 + m_{P}^2 - m_{l}^2)^2 - 4 m_{\tau}^2 m_{P}^2}
/ (2 m_{\tau} )$.
In the above expression, the decay constant $f_P$ is given by $f_P = f_{\pi}$ with $a_P = 1$ for
$\tau^- \to \pi^0 l$, and $f_P = f^s_{\eta^{(\prime)}}$ with $a_P = 1/2$ for $\tau^- \to\eta^{(\prime)} l$.
In the case of $\tau^- \to\eta^{(\prime)} l$, the $u$ and $d$ quark contributions to
the matrix element $\langle \eta^{(\prime)} | {\bar q} \gamma_{\alpha} \gamma_5 q | 0 \rangle$
cancel each other in Eq.~(\ref{Tau-decay-amp}) so that only the $s$ quark contribution to
the decay constant, $f^s_{\eta^{(\prime)}}$, remains.

Similarly, the decay rate for $\tau^- \to V l$ $(P = \rho^0, \omega, \phi$, and $l = e,~ \mu)$ is
given by
\begin{eqnarray}
\Gamma = a_V {G_F^2 f_{V}^2 m_V^2 \over \pi m_{\tau}^2} ~| \epsilon_{l
\tau} |^2 \left| {\vec p}_{_{l}} \right| \left[ m_{\tau}^2 + m_l^2 -m_V^2
+ {1 \over m_V^2} (m_{\tau}^2 + m_V^2 -m_l^2) (m_{\tau}^2 - m_V^2 -m_l^2) \right],
\end{eqnarray}
where $|\vec p_{_l}| = \sqrt{(m_{\tau}^2 + m_{V}^2 - m_{l}^2)^2 - 4 m_{\tau}^2 m_{V}^2}
/ (2 m_{\tau} )$.
The decay constant $f_V$ is given by $f_V = f_{\rho}$ with $a_V = (1/2 -\sin^2 \theta_W)^2$ for
$\tau^- \to \rho^0 l$, and $f_V = f_{\omega}$ with $a_V = (\sin^2 \theta_W /3)^2$ for
$\tau^- \to \omega l$, and $f_V = f_{\phi}$ with $a_V = 2 (1/4 -\sin^2 \theta_W /3)^2$ for
$\tau^- \to \phi l$.

Using the current experimental bounds on the branching ratios, we find the constraints on the
parameters $|\epsilon_{e \tau}|$ and $|\epsilon_{\mu \tau}|$ which are shown in Table~\ref{table:1}.
Notice that the constraint on $|\epsilon_{e \tau}|$ obtained from $\tau^- \to \pi^0 e^-$ is
$|\epsilon_{e \tau}| < 4.2 \times 10^{-4}$, which is more stringent than the so far most stringent
bound obtained from $\tau \to e \bar e e$ as shown in Table~\ref{table:4}.
The constraints on $|\epsilon_{\mu \tau}|$ obtained from $\tau^- \to \pi^0 \mu^-$ and
$\tau^- \to \rho^0 \mu^-$
are comparable to the so far most stringent bound shown in Table~\ref{table:4}.
The upper bounds on $|\epsilon_{e (\mu) \tau}|$ from $\tau^- \to \eta^{(\prime)} l$ and
$\tau^- \to \omega l$ are weaker.
\\

\begin{table}
\caption{Constraints from $\tau \to P l$ and $\tau \to V l$.}
\smallskip
\begin{tabular}{|c|c|c|}
\hline
  Process & Branching Ratio & Constraint on $|\epsilon_{_{l l^{\prime}}}|$  \\
\hline
$\tau^- \to \pi^0 e^-$ & $< 8.0 \times 10^{-8}$ & $|\epsilon_{e \tau}| < 4.2 \times 10^{-4}$ \\
$\tau^- \to \pi^0 \mu^-$ & $< 1.1 \times 10^{-7}$ & $|\epsilon_{\mu \tau}| < 7.0 \times 10^{-4}$ \\
$\tau^- \to \eta e^-$ & $< 9.2 \times 10^{-8}$ & $|\epsilon_{e \tau}| < 1.2 \times 10^{-3}$ \\
$\tau^- \to \eta \mu^-$ & $< 6.5 \times 10^{-8}$ & $|\epsilon_{\mu \tau}| < 9.7 \times 10^{-4}$ \\
$\tau^- \to \eta^{\prime} e^-$ & $< 1.6 \times 10^{-7}$ & $|\epsilon_{e \tau}| < 1.0 \times 10^{-3}$ \\
$\tau^- \to \eta^{\prime} \mu^-$ & $< 1.3 \times 10^{-7}$ & $|\epsilon_{\mu \tau}| < 1.0 \times 10^{-3}$ \\
\hline
$\tau^- \to \rho^0 e^-$ & $< 6.3 \times 10^{-8}$ & $|\epsilon_{e \tau}| < 6.5 \times 10^{-4}$ \\
$\tau^- \to \rho^0 \mu^-$ & $< 6.8 \times 10^{-8}$ & $|\epsilon_{\mu \tau}| < 6.8 \times 10^{-4}$ \\
$\tau^- \to \omega e^-$ & $< 1.1 \times 10^{-7}$ & $|\epsilon_{e \tau}| < 3.2 \times 10^{-3}$ \\
$\tau^- \to \omega \mu^-$ & $< 8.9 \times 10^{-8}$ & $|\epsilon_{\mu \tau}| < 2.5 \times 10^{-3}$ \\
$\tau^- \to \phi e^-$ & $< 7.3 \times 10^{-8}$ & $|\epsilon_{e \tau}| < 7.5 \times 10^{-4}$ \\
$\tau^- \to \phi \mu^-$ & $< 1.3 \times 10^{-7}$ & $|\epsilon_{\mu \tau}| < 1.0 \times 10^{-3}$ \\
\hline
\end{tabular}
\label{table:1}
\end{table}


\noindent
{\bf Constraints from $V \to l\bar l'$ and $P \to l \bar l'$}

Here $V$ can be a vector meson $J/\psi$ or $\Upsilon$, and $P$ can be a pseudoscalar meson
$\pi^0$, $\eta$ or $\eta'$.  The $l$ and $l'$ stand for charged leptons with different flavors
$l \neq l'$.
These processes can be induced by exchange $Z$ boson between quarks and leptons.
The general decay amplitude for $M \to l \bar l'$ (where $M$ denotes either $V$ or $P$)
is given by
\begin{eqnarray}
\label{2body-Meson-decay-amp}
{\cal M} &=& 2\sqrt{2} G_F ~\epsilon_{l l^{\prime}} \sum_{q = u, d, s, c, b}
\langle 0 | \bar q \gamma_{\alpha} (I_3 P_L - Q_q \sin^2\theta_W) q | M (p_{_M}) \rangle
\cdot \left[ \bar u_{l}(p_1) \gamma^{\alpha} (1 - \gamma_5) v_{l^{\prime}}(p_2) \right]
\nonumber\\
 &=& 2\sqrt{2} G_F ~\epsilon_{l l^{\prime}} \sum_{q = u, d, s, c, b}
\langle 0 | \bar q \gamma_{\alpha} (g^q_{_V} + g^q_{_A} \gamma_5 ) q | M (p_{_M}) \rangle
\cdot \left[ \bar u_{l}(p_1) \gamma^{\alpha} (1 - \gamma_5) v_{l^{\prime}}(p_2) \right],
\end{eqnarray}
where we use the decay constants $f_{J/\Psi} = 416~{\rm MeV}$ and
$f_{\Upsilon (3S)} = 430~{\rm MeV}$~\cite{pdg,fV}.
Again, exchange of Higgs boson does not contribute to these two classes of processes since
$\langle 0\vert \bar q q\vert M\rangle = 0$.

The decay rate for $V \to l \bar l^{\prime}$ $(V = J/\Psi, \Upsilon)$ is found to be
\begin{eqnarray}
\label{2body-Vector-decay-rate}
\Gamma = {8 G_F^2 f_{V}^2 \over 3 \pi} (g^q_{_V})^2 |\epsilon_{l l^{\prime}} |^2 ~|\vec p_{_l}|
\left[ m_{V}^2 - {1 \over 2} m_{l} - {1 \over 2} m_{l^{\prime}} -{1 \over 2 m_{V}^2}
(m_{l}^2 -m_{l^{\prime}}^2)^2 \right],
\end{eqnarray}
where $|\vec p_{_l}| = \sqrt{(m_{V}^2 + m_{l}^2 - m_{l^{\prime}}^2)^2 - 4 m_{V}^2 m_{l}^2}
/ (2 m_V )$, and $g^q_{_V} = g^c_{_V}$ for $V = J/\Psi$ and $g^q_{_V} = g^b_{_V}$ for
$V = \Upsilon$.

Similarly the rate of a pseudoscalar meson decay $P \to l \bar l^{\prime}$
$(P = \pi^0, \eta, \eta^{\prime})$ is given by
\begin{eqnarray}
\label{2body-Pseudoscalar-decay-rate}
\Gamma = a_P {G_F^2 f_{P}^2 \over 2 \pi m_P} |\epsilon_{l l^{\prime}} |^2 ~|\vec p_{_l}|
\left[ (m_l^2 +m_{l^{\prime}}^2) m_P^2 -(m_l^2 -m_{l^{\prime}}^2)^2 \right],
\end{eqnarray}
where $|\vec p_{_l}| = \sqrt{(m_{P}^2 + m_{l}^2 - m_{l^{\prime}}^2)^2 - 4 m_{P}^2 m_{l}^2}
/ (2 m_P )$, and $a_P = 1$, $f_P = f_{\pi}$ for $P = \pi^0$, and $a_P = 1/2$,
$f_P = f^s_{\eta^{(\prime)}}$ for $P = \eta^{(\prime)}$.
Note that as in the case of $\tau^- \to\eta^{(\prime)} l$, only the $s$ quark contribution to
the decay constant, $f^s_{\eta^{(\prime)}}$, appears in $\eta^{(\prime)} \to l ~\bar l^{\prime}$.
We find that the constraints on $|\epsilon_{l l^{\prime}}|$ from these two body meson decays are
rather weak as summarized in Table~\ref{table:2}. The constraints obtained are much weaker than those
obtained in the previous section.
\\

\begin{table}
\caption{Constraints from $V \to l \bar l'$ and $P \to l \bar l'$.}
\smallskip
\begin{tabular}{|c|c|c|}
\hline
  Process & Branching Ratio & Constraint on $|\epsilon_{_{l l^{\prime}}}|$  \\
\hline
$\Upsilon (3S) \to e^{\pm} \tau^{\mp}$ & $< 5 \times 10^{-6}$ & $|\epsilon_{e \tau}| < 0.39$ \\
$\Upsilon (3S) \to \mu^{\pm} \tau^{\mp}$ & $< 4.1 \times 10^{-6}$ & $|\epsilon_{\mu \tau}| < 0.35$ \\
$J/\Psi (1S) \to e^{\pm} \mu^{\mp}$ & $< 1.1 \times 10^{-6}$ & $|\epsilon_{e \mu}| \sim O(1)$ \\
$J/\Psi (1S) \to e^{\pm} \tau^{\mp}$ & $< 8.3 \times 10^{-6}$ & $|\epsilon_{e \tau}| \sim O(1)$ \\
$J/\Psi (1S) \to \mu^{\pm} \tau^{\mp}$ & $< 2.0 \times 10^{-6}$ & $|\epsilon_{\mu \tau}| \sim O(1)$ \\
\hline
$\pi^0 \to e^+ \mu^-$ & $< 3.4 \times 10^{-9}$ & $|\epsilon_{e \mu}| < 0.80$ \\
$\pi^0 \to e^- \mu^+$ & $< 3.8 \times 10^{-10}$ & $|\epsilon_{e \mu}| < 0.27$ \\
$\eta \to e^{\pm} \mu^{\mp}$ & $< 6 \times 10^{-6}$ & $|\epsilon_{e \mu}| \sim O(1)$ \\
$\eta^{\prime} \to e^{\pm} \mu^{\mp}$ & $< 4.7 \times 10^{-4}$ & $|\epsilon_{e \mu}| \sim O(1)$ \\
\hline
\end{tabular}
\label{table:2}
\end{table}


\noindent
{\bf Constraints from $M \to M' l\bar l'$}

We now consider semileptonic three body decays of the type $M \to M^{\prime} l \bar l'$
with $M = B, K$ and $M^{\prime} = K, K^*, \pi$, such as
$B \to K l \bar l^{\prime}$, $B \to K^* l \bar l^{\prime}$, $B \to \pi l \bar l^{\prime}$,
and $K \to \pi l \bar l^{\prime}$.
These decays can occur through quark level subprocesses $b \to s l \bar l^{\prime}$
or $s \to d l \bar l^{\prime}$.  The FCNC $b \to s$ or $s \to d$ transition can arise via
$Z$-penguin and Higgs-penguin diagrams at one loop level the same way as in the SM.
After taking into account the SM effective $b$-$s$-$Z$ and $b$-$s$-Higgs couplings (or
$s$-$d$-$Z$ and $s$-$d$-Higgs couplings)~\cite{Z_penguin,Higgs_penguin}, the lepton flavor violating
FCNC processes $b \to s l \bar l^{\prime}$ (or $s \to d l \bar l^{\prime}$) can occur at tree level
via the couplings given in Eq.~(\ref{Heta}).

The decay amplitude for $M \to M^{\prime} l \bar l^{\prime}$ $[{\rm where}~M = B ~({\rm or}~ K);
~ M^{\prime} = K, K^*, \pi ~({\rm or}~ \pi)$; $l,~l^{\prime} = e, ~\mu, ~\tau ~(l \neq l^{\prime}))]$
is given by
\begin{eqnarray}
\label{3body-decay-amp}
{\cal M} = {\cal M}^{Z} +{\cal M}^{h},
\end{eqnarray}
where ${\cal M}^{Z}$ and ${\cal M}^{h}$ denote the $Z$-mediated and Higgs-mediated decay amplitude,
respectively, in the following form
\begin{eqnarray}
\label{3body-decay-amp-Z}
{\cal M}^Z &=& -{1 \over 32 \pi^2} V_{i q^{\prime \prime}}^* V_{i q^{\prime}}
{g^4 \over \cos^2 \theta_W M_W^2} ~C_0 (x_i) ~\epsilon_{l l^{\prime}}
~\langle M^{\prime} (p^{\prime}) | \bar q^{\prime \prime} \gamma_{\alpha} (1 -\gamma_5) q^{\prime}
| M(p) \rangle  \nonumber \\
&& \times \left[ \bar u_{l}(k_1) \gamma^{\alpha} (1 - \gamma_5) v_{l^{\prime}}(k_2) \right] ,
\end{eqnarray}
\begin{eqnarray}
\label{3body-decay-amp-Higgs}
{\cal M}^h &=& i {9 \over 1024 \pi^2} V_{t q^{\prime \prime}}^* V_{t q^{\prime}}
g^4 {m_t^2 m_{q^{\prime}} \over m_W^4 m_h^2} ~\epsilon_{l l^{\prime}}
~\langle M^{\prime} (p^{\prime}) | \bar q^{\prime \prime} (1 +\gamma_5) q^{\prime}
| M(p) \rangle  \nonumber \\
&& \times \left\{ \bar u_{l}(k_1) [ ( m_l +m_{l^{\prime}}) +(m_{l^{\prime}} -m_l) \gamma_5 ]
v_{l^{\prime}}(k_2) \right\} ,
\end{eqnarray}
where (i) for $B \to K^{(*)} l \bar l^{\prime}$, $q^{\prime} = b$ and $q^{\prime \prime} = s$,
(ii) for $B \to \pi l \bar l^{\prime}$, $q^{\prime} = b$ and $q^{\prime \prime} = d$,
(iii) for $K \to \pi l \bar l^{\prime}$, $q^{\prime} = s$ and $q^{\prime \prime} = d$.
The $V_{i q^{\prime}}$ denotes the CKM matrix element with $i = t, c, u$ and
$C_0 (x_i) = (x_i / 8) \left[ {(x_i -6) /(x_i -1)} + {(3 x_i +2) \ln x_i / (x_i -1)^2}
\right]$ with $x_i = {m_i^2 / m_W^2}$~\cite{Z_penguin}.

Compared with the Z-mediated amplitude, the Higgs-mediated amplitude is negligibly small, since
$m_h \gg m_b , m_l$, so
that the Higgs contribution can be safely neglected.  For example, in the cases
of $B \to K^{(*)} l \bar l^{\prime}$ and $K \to \pi l \bar l^{\prime}$ decays,
$\left| {\cal M}^h /{\cal M}^Z \right|$ is suppressed roughly by $O(x_t (m_b m_l / m_h^2))$
and $O(x_t (m_s m_l / m_h^2))$, respectively.

For $B \to P l \bar l^{\prime}$ $(P = \pi, K)$, the form factors $F_1$ and
$F_0$ (or $f_+$ and $f_-$) are defined by
\begin{eqnarray}
\label{BK-formfactor}
\langle P (p^{\prime}) | \bar s \gamma_{\alpha} (1 -\gamma_5) b | B (p) \rangle
&=& F_1 (q^2) \left[ (p + p^{\prime})_{\alpha} - {m_B^2 -m_K^2 \over q^2} q_{\alpha} \right]
    + F_0 (q^2) {m_B^2 -m_K^2 \over q^2} q_{\alpha}  \nonumber \\
&=& f_+ (q^2) (p + p^{\prime})_{\alpha} + f_- (q^2) q_{\alpha},
\end{eqnarray}
where $q \equiv p - p^{\prime}$.
For $B \to K^* l \bar l^{\prime}$, the form factors $V$, $A_0$, $A_1$, and $A_2$ are defined by
\begin{eqnarray}
\label{BKstar-formfactor}
\langle K^* (p^{\prime}, \epsilon) | \bar s \gamma_{\alpha} (1 -\gamma_5) b | B (p) \rangle
&=& -\epsilon_{\alpha \beta \rho \sigma} \epsilon^{\beta *} p^{\rho} p^{\prime \sigma}
 {2 V(q^2) \over m_B + m_{K^*}} \nonumber \\
&& -i \left( \epsilon_{\alpha}^* - {\epsilon^* \cdot q \over q^2} q_{\alpha} \right) (m_B + m_{K^*}) A_1 (q^2)
\nonumber \\
&& +i \left( (p + p^{\prime})_{\alpha} - {m_B^2 -m_{K^*}^2 \over q^2} q_{\alpha} \right) (\epsilon^* \cdot q)
 {A_2 (q^2) \over m_B + m_{K^*}}  \nonumber \\
&& -i {2 m_{K^*} (\epsilon^* \cdot q) \over q^2} q_{\alpha} A_0 (q^2) ,
\end{eqnarray}
where $\epsilon$ is the polarization vector of the $K^*$ meson.
For numerical analysis, we use the form factors calculated in the framework of light-cone QCD sum
rules~\cite{BK-formfactor}.  The $q^2$ dependence of the form factors can be expressed as
\begin{eqnarray}
\label{formfactor}
F(q^2) = {F(0) \over 1 - a_F {q^2 \over m_B^2} + b_F \left( {q^2 \over m_B^2} \right)^2 }~,
\end{eqnarray}
where the values of the parameters $F(0)$, $a_F$ and $b_F$ for $B \to \pi$, $B \to K$ and $B \to K^*$ are
given in~\cite{BK-formfactor}.

Summing over the spins of the final leptons, we obtain
\begin{eqnarray}
\label{decay_rate-BtoKll}
{d \Gamma (B \to P l \bar l^{\prime}) \over d q^2}
&=& {1 \over 192 \pi^5} {G_F^2 \alpha^2 \over \sin^4 \theta_W \cos^4 \theta_W}
\left| V^*_{ts} V_{tb} \right|^2 C^2_0 (x_t) |\epsilon_{l l^{\prime}}|^2
~{\lambda^{3/2}(m_B^2, m_P^2, q^2) \over m_B^3}  \nonumber \\
&& \times (1 -2 \rho)^2 \left[ (1 +\rho) \left| f_+ (q^2) \right|^2 +3 \rho \left| f_0 (q^2) \right|^2 \right],
\end{eqnarray}
where $\lambda(a, b, c) = (a-b-c)^2 -4bc$, $\rho = m_l / (2 q^2)$ and
\begin{eqnarray}
\label{f0_rho}
f_0 (q^2) \equiv {(m_B^2 -m_P^2) f_+ (q^2) + q^2 f_- (q^2) \over \lambda^{1/2} (m_B^2, m_P^2, q^2)}.
\end{eqnarray}
Here the mass of only one light lepton in the final state has been neglected so that the parameter $\rho$
represents the effect of the remaining lepton mass, e.g. $m_{\tau}$.  Thus, for $B \to K e \mu$ decays,
$\rho$ can be neglected.
The decay rate for $B \to K^* l \bar l^{\prime}$, summed over the spins of the final leptons and $K^*$,
is given by
\begin{eqnarray}
\label{decay_rate-BtoKstarll}
{d \Gamma (B \to K^* l \bar l^{\prime}) \over d s}
&=& {1 \over 768 \pi^5} {G_F^2 \alpha^2 \over \sin^4 \theta_W \cos^4 \theta_W}
 \left| V^*_{ts} V_{tb} \right|^2 C^2_0 (x_t) |\epsilon_{l l^{\prime}}|^2
 ~ m_B^3 \tilde \lambda^{1/2}  \nonumber \\
&& \times \left\{ \left| V(q^2) \right|^2 {8 m_B^4 s \tilde \lambda \over (m_B +m_{K^*})^2} \right.
 \nonumber \\
&& ~~ + \left| A_1 (q^2) \right|^2 (m_B +m_{K^*})^2 \left( {\tilde \lambda \over r} + 12 s \right)
 \nonumber \\
&& ~~ + \left| A_2 (q^2) \right|^2 {m_B^4 \over (m_B +m_{K^*})^2} {\tilde \lambda^2 \over r} \nonumber \\
&& ~~ \left. - 2 m_B^2 ~{\rm Re}\left[ A_1 (q^2) A_2^* (q^2) \right] {\tilde \lambda (1 -r -s) \over r}
\right\},
\end{eqnarray}
where $r = m_{K^*}^2 / m_B^2$, $s = q^2 / m_B^2$, and $\tilde \lambda = 1 +r^2 +s^2 -2r -2s -2rs$.
The branching ratios for $B \to \pi l \bar l^{\prime}$, $B \to K l \bar l^{\prime}$ and
$B \to K^* l \bar l^{\prime}$ can be calculated after the decay rates given in
Eqs.~(\ref{decay_rate-BtoKll}) and (\ref{decay_rate-BtoKstarll}) are integrated in the range
$(m_l + m_{l^{\prime}})^2 \leq q^2 \leq (m_B -m_{M^{\prime}})^2$.
From the current experimental bounds on those branching ratios, we obtain the constraints on
$\epsilon_{ll'}$ shown in Table~\ref{table:3}.

For $K \to \pi l \bar l'$, we normalize the branching ratio to $K^+ \to \pi^0 e^+ \nu_e$ and neglect
the phase factor difference~\cite{desh}. We have
\begin{eqnarray}
\label{K_pill}
&&{B(K^+\to \pi^+ l \bar l')\over B(K^+\to \pi^0 e^+ \nu_e)}
= {2 \alpha^2\over \pi^2 \sin^4 \theta_W \cos^4 \theta_W}
\left \vert {V_{ts}^*V_{td}\over V_{us}}\right \vert^2 C^2_0(x_t) ~\vert \epsilon_{ll'}\vert^2
\;,\nonumber\\
&&{B(K_L\to \pi^0 l \bar l')\over B(K^+\to \pi^0 e^+ \nu_e)}
= {\tau_{K_L}\over \tau_{K^+}}{2 \alpha^2\over \pi^2 \sin^4 \theta_W \cos^4 \theta_W}
\left \vert {\rm Im} \left({V_{ts}^*V_{td} \over V_{us}} \right) \right \vert^2
C^2_0(x_t) ~\vert \epsilon_{ll'}\vert^2\;,
\end{eqnarray}
where $\tau_K$ is the lifetime of the Kaon.  Note that the model-dependent form factors do not appear
in the above formulas.
Using the experimental value $B(K^+ \to \pi^0 e^+ \nu_e) = (5.08 \pm 0.05) \%$~\cite{pdg}, we obtain
the constraints on $\epsilon_{ll'}$ shown in Table~\ref{table:3}.
Alternatively, the decay rate for $K \to \pi l \bar l'$ can be calculated by using
Eq.~(\ref{decay_rate-BtoKll}).  In this case, the mass of muon is not neglected and the parameter
$\rho = m_{\mu} / (2 q^2)$.
The relevant form factors are given by
\begin{eqnarray}
\label{formfactor-Kpi}
&& f_+^{K \pi} (q^2) \simeq -1 - \lambda_+ q^2 ~, \nonumber \\
&& \tilde f_0^{K \pi} (q^2) \equiv f_+^{K \pi} (q^2) + {q^2 \over m_K^2 - m_{\pi}2} f_-^{K \pi} (q^2)
 \simeq -1 - \lambda_0 q^2 ~,
\end{eqnarray}
where $\lambda_+ = 0.067 ~{\rm fm^2}$ and $\lambda_0 = 0.040 ~{\rm fm^2}$~\cite{Donoghue}.
The constraints on $\epsilon_{ll'}$ obtained in this way (number shown in the bracket for
$K^+ \to \pi^+ e^+ \mu^-$) is similar to those obtained by using Eq.~(\ref{K_pill}) as shown in
Table~\ref{table:3}. The constraints obtained here are again much weaker than those obtained from
$\tau \to P l$.
\\

\begin{table}
\caption{Constraints from $M \to M^{\prime} l \bar l'$.}
\smallskip
\begin{tabular}{|c|c|c|}
\hline
  Process & Branching Ratio & Constraint on $|\epsilon_{_{l l^{\prime}}}|$  \\
\hline
$B^+ \to \pi^+ e^+ \mu^-$ & $< 6.4 \times 10^{-3}$ & $|\epsilon_{e \mu}| \sim O(1)$ \\
$B^+ \to \pi^+ e^- \mu^+$ & $< 6.4 \times 10^{-3}$ & $|\epsilon_{e \mu}| \sim O(1)$ \\
$B^+ \to \pi^+ e^{\pm} \mu^{\mp}$ & $< 1.7 \times 10^{-7}$ & $|\epsilon_{e \mu}| < 0.56$ \\
$B^+ \to K^+ e^+ \mu^-$ & $< 9.1 \times 10^{-8}$ & $|\epsilon_{e \mu}| < 0.18$ \\
$B^+ \to K^+ e^- \mu^+$ & $< 1.3 \times 10^{-7}$ & $|\epsilon_{e \mu}| < 0.21$ \\
$B^+ \to K^+ e^{\pm} \mu^{\mp}$ & $< 9.1 \times 10^{-8}$ & $|\epsilon_{e \mu}| < 0.12$ \\
$B^+ \to K^+ \mu^{\pm} \tau^{\mp}$ & $< 7.7 \times 10^{-5}$ & $|\epsilon_{\mu \tau}| \sim O(1)$ \\
$B^0 \to \pi^0 e^{\pm} \mu^{\mp}$ & $< 1.4 \times 10^{-7}$ & $|\epsilon_{e \mu}| < 0.73$ \\
$B^0 \to K^0 e^{\pm} \mu^{\mp}$ & $< 2.7 \times 10^{-7}$ & $|\epsilon_{e \mu}| < 0.21$ \\
\hline
$B^+ \to K^* (892)^+ e^+ \mu^-$ & $< 1.3 \times 10^{-6}$ & $|\epsilon_{e \mu}| < 7.1 \times 10^{-2}$ \\
$B^+ \to K^* (892)^+ e^- \mu^+$ & $< 9.9 \times 10^{-7}$ & $|\epsilon_{e \mu}| < 6.2 \times 10^{-2}$ \\
$B^+ \to K^* (892)^+ e^{\pm} \mu^{\mp}$ & $< 1.4 \times 10^{-7}$ & $|\epsilon_{e \mu}| < 1.7 \times 10^{-2}$ \\
$B^0 \to K^* (892)^0 e^+ \mu^-$ & $< 5.3 \times 10^{-7}$ & $|\epsilon_{e \mu}| < 4.5 \times 10^{-2}$ \\
$B^0 \to K^* (892)^0 e^- \mu^+$ & $< 3.4 \times 10^{-7}$ & $|\epsilon_{e \mu}| < 3.6 \times 10^{-2}$ \\
$B^0 \to K^* (892)^0 e^{\pm} \mu^{\mp}$ & $< 5.8 \times 10^{-7}$ & $|\epsilon_{e \mu}| < 3.4 \times 10^{-2}$ \\
\hline
$K^+ \to \pi^+ e^+ \mu^-$ & $< 1.3 \times 10^{-11}$ & $|\epsilon_{e \mu}| < 0.44 ~[0.8]$ \\
$K^+ \to \pi^+ e^- \mu^+$ & $< 5.2 \times 10^{-10}$ & $|\epsilon_{e \mu}| \sim O(1)$ \\
$K_L \to \pi^0 e^{\pm} \mu^{\mp}$ & $< 6.2 \times 10^{-9}$ & $|\epsilon_{e \mu}| \sim O(1)$ \\
\hline
\end{tabular}
\label{table:3}
\end{table}

\begin{table}
\caption{Constraints from $l_i \to l_j \bar l_k l_l$,
$l_i \to l_j \gamma$ decays and $\mu - e$ conversion.}
\smallskip
\begin{tabular}{|c|c|c|}
\hline
  Process & Conversion rate & Constraint on $|\epsilon_{_{l l^{\prime}}}|$  \\
\hline
$\mu - e$ conversion & $< 4.3 \times 10^{-12}$ & $|\epsilon_{e \mu}| < 1.7 \times 10^{-7}$ \\
\hline
  Process & Branching Ratio & Constraint on $|\epsilon_{_{l l^{\prime}}}|$  \\
  \hline
$\mu^- \to e^+ e^- e^-$ & $< 1 \times 10^{-12}$ & $|\epsilon_{e \mu}| < 1.1 \times 10^{-6}$ \\
$\tau^- \to e^+ e^- e^-$ & $< 3.6 \times 10^{-8}$ & $|\epsilon_{e \tau}| < 5.1 \times 10^{-4}$ \\
$\tau^- \to \mu^+ \mu^- \mu^-$ & $< 3.2 \times 10^{-8}$ & $|\epsilon_{\mu \tau}| < 4.9 \times 10^{-4}$ \\
$\tau^- \to \mu^+ \mu^- e^-$ & $< 4.1 \times 10^{-8}$ & $|\epsilon_{e \tau}| < 7.2 \times 10^{-4}$ \\
$\tau^- \to e^+ e^- \mu^-$ & $< 2.7 \times 10^{-8}$ & $|\epsilon_{\mu \tau}| < 5.6 \times 10^{-4}$ \\
$\mu^- \to e \gamma$ & $< 1 \times 10^{-15}$ & $|\epsilon_{e \mu}| \lesssim 1.1 \times 10^{-4}$ \\
$\tau^- \to e \gamma$ & $< 5 \times 10^{-11}$ & $|\epsilon_{e \tau}| \lesssim 2.4 \times 10^{-2}$ \\
$\tau^- \to \mu \gamma$ & $< 4 \times 10^{-11}$ & $|\epsilon_{\mu \tau}| \lesssim 1.5 \times 10^{-2}$ \\
\hline
\end{tabular}
\label{table:4}
\end{table}

\begin{table}
\caption{Constraints on $\epsilon_{ll'}$ from $Z\to l \bar l'$ decays.}
\smallskip
\begin{tabular}{|c|c|c|}
\hline
  Process & Branching Ratio & Constraint on $|\epsilon_{_{l l^{\prime}}}|$  \\
\hline
$Z \to e^{\pm} \mu^{\mp}$ & $< 1.7 \times 10^{-6}$ & $|\epsilon_{e \mu}| < 1.8 \times 10^{-3}$ \\
$Z \to e^{\pm} \tau^{\mp}$ & $< 9.8 \times 10^{-6}$ & $|\epsilon_{e \tau}| < 4.3 \times 10^{-3}$ \\
$Z \to \mu^{\pm} \tau^{\mp}$ & $< 1.2 \times 10^{-5}$ & $|\epsilon_{\mu \tau}| < 4.8 \times 10^{-3}$ \\
\hline
\end{tabular}
\label{table:5}
\end{table}


\noindent
{\bf Constraint from muonium-antimuonium oscillation}

At tree level, exchange of $Z$ boson can generate an effective Hamiltonian
of the form
\begin{eqnarray}
H_{eff} = \sqrt{2} G_F \epsilon_{e\mu}^{*2} \bar \mu \gamma_\mu (1-\gamma_5)
e \bar \mu \gamma^\mu (1-\gamma_5) e\;.
\end{eqnarray}
This interaction will result in muonium-antimuonium oscillation.

The SM prediction for muonium and antimuonium oscillation is
extremely small. Observation of this oscillation at a
substantially larger rate will be an indication of new physics.
Experimentally, no oscillation has been observed. The current
upper limit for the probability of spontaneous muonium to
antimuonium conversion was established at $P_{\bar MM}\leq 8.3\times 10^{-11}$
(90\% C.L.) in 0.1 T magnetic field~\cite{Willmann:1998gd}.

In the absence of external electromagnetic fields, the probability
$P_{\bar MM}$ of observing a transition  can be written
as~\cite{formula} $P_{\bar MM}(0 \mbox{T})\simeq
{|\delta|^2}/{(2\Gamma_\mu^2)}$, where $\delta\equiv 2\langle\bar
M|H_{eff}|M\rangle$ and $\Gamma_\mu$ is the muon decay width. For $H_{eff}$ given above,
the transition amplitude
is given by $\delta=32 G_{F}\epsilon^2_{e\mu}/(\sqrt{2}\pi a^3)$  for both
triplet and singlet muonium states, where $a\simeq (\alpha
m_e)^{-1}$ is the Bohr radius. The probability $P_{\bar MM}$ has
strong magnetic field dependence which usually occurs in
experimental situation. With an external magnetic field, there is
a reduction factor $S_B$, i.e. $P_{\bar MM}(B)=S_B P_{\bar MM}(0
\mbox{T})$. The magnetic field correction factor $S_B$ describes
the suppression of the conversion in the external magnetic field
due to the removal of degeneracy between corresponding levels in
$\bar M$ and $M$. One has $S_B=0.35$ for our case at
$B=0.1 \mbox{T}$~\cite{Willmann:1998gd,Horikawa:1995ae}.
Using this experimental information, one obtains a constraint
\begin{eqnarray}
\vert \epsilon_{e\mu}\vert < 4\times 10^{-2}.
\end{eqnarray}
This constraint is rather weak compared with that from $\mu - e$ conversion.

Exchange of Higgs boson will also contribute. But this contribution is suppressed by a factor
$m^2_\mu/m_h^2$ and can be safely neglected compared with $Z$ boson exchange contribution.
\\


\noindent
{\bf Constraints from $l_i \to l_j \bar l_k l_l$,
$l_i \to l_j \gamma$, $Z\to l \bar l'$  decays and $\mu - e$ conversion}

These processes have been studied in the literature before~\cite{fcnc, fcnc3}. For comparison,
we summarize the results for constraints on $\epsilon_{ll'}$ for
$l_i \to l_j \bar l_k l_l$, $l_i \to l_j \gamma$ and $\mu -e $ conversion, and
$Z\to l \bar l'$~\cite{footnote}
in Tables~\ref{table:4} and \ref{table:5}, respectively.
The most stringent upper bound
on $|\epsilon_{e \mu}|$ is of order $10^{-7}$ from $\mu - e$ conversion
in atomic nuclei.  The upper bounds on $|\epsilon_{e \tau}|$ and
$|\epsilon_{\mu \tau}|$ obtained are of order $10^{-4}$ from
$\tau \to e \bar e e$ and $\tau \to \mu \bar \mu \mu$ decays.
\\


\noindent
{\bf Discussions on the mixing matrix $U_{\nu N}$ between the light and heavy neutrinos}

We now discuss some implications of the constraints obtained earlier on the model parameters.
In this model, to the order we are studying, the light neutrino mass is related to $U_{\nu N}$ with
\begin{eqnarray} \label{mnu_Msigma}
U_{\rm PMNS} \hat m_\nu U_{\rm PMNS}^T =- U_{\nu N}^{}M_{\Sigma} U_{\nu N}^T \,\,,
\end{eqnarray}
where the light neutrino mass matrix $\hat m_{\nu}$ is diagonal:
\begin{eqnarray}
\hat m_\nu = {\rm diag}\left( m_{\nu_1},m_{\nu_2},m_{\nu_3}
\right) = U_{\rm PMNS}^\dagger m_\nu^{} U_{\rm PMNS}^* \,\,.
\end{eqnarray}
Thus, one might think that the elements of $U_{\nu N}$ are too small to be relevant to the FCNC
discussion, because with only one generation of the light and heavy neutrinos, $|U_{\nu N}|$ is simply
given by $(m_\nu/M_\Sigma)^{1/2}$. It leads to the fact that for $M_{\Sigma} > 100$ GeV,
$U_{\nu N}$ is less than $10^{-5}$, since the light neutrino masses must be less than an eV or so.
If with more than one generation of the light and heavy neutrinos, all elements of $U_{\nu N}$ are
the same order of magnitudes (the canonical seesaw models), the resulting elements of the $\epsilon$
matrix will all be way below the constraints we have obtained.  This makes the model irrelevant for an
experimental detection. The FCNC study of the kind studied here is therefore not interesting for
canonical seesaw models.
However, it has been shown that with more than one generation of the light and heavy neutrinos, there are
non-trivial solutions of $U_{\nu N}$ such that the right hand side of Eq.~(\ref{mnu_Msigma}) becomes
exactly zero but the elements of $U_{\nu N}$ can be arbitrarily large~\cite{large_mixing,ours}.
Thus, these solutions evade the canonical seesaw constraint $|U_{\nu N}| = (m_\nu/M_\Sigma)^{1/2}$
held in the one generation case~\cite{large_mixing,ours}.
It is interesting if one can find the $U_{\nu N}$ which satisfies
existing experimental constraints by adding small perturbations to the above non-trivial solutions.
A recent study has shown such solutions of $U_{\nu N}$ that indeed can have large elements and
satisfy the current experimental constraints~\cite{ours}.
In the following we will describe some of those solutions having relevance to our FCNC study.

Let us indicate the solution of $U_{\nu N}$ which gives zero light neutrino mass as $U_0$. We then
add a perturbation $U_\delta$ to $U_{0}$ such that $U_{\nu N} = U_0 + U_\delta$.
Since $U_0 M_{\Sigma} U_0^T = 0$, the neutrino mass matrix is given by
\begin{eqnarray}
m_\nu = - U_0 M_{\Sigma} U_\delta^T - U_\delta M_{\Sigma} U^T_0 - U_\delta M_{\Sigma} U^T_\delta ~.
\end{eqnarray}
If the first two terms are not zero, the matrix elements $a_{ij}$ in $U_0$ and $\delta_{ij}$ in
$U_\delta$ are of order $a_{ij} \delta_{ij} \sim m_\nu / M_{\Sigma}$ which is much smaller than 1.
Since we are interested in having large $a_{ij}$, the elements $\delta_{ij}$ must be much smaller
than $a_{ij}$, and the third term, for practical purpose, can be neglected. If on the other hand,
the first two terms are zero, the third term must be kept.
The elements of $U_\delta$ in this case are of order $(m_\nu/M_\Sigma)^{1/2}$.

In the basis where $M_{\Sigma}$ is diagonal, one can write
\begin{eqnarray}
M_{\Sigma} \,\,=\,\, \hat M_{\Sigma} \,\,=\,\,
{\rm diag}\Biggl(\frac{1}{r_1},\,\frac{1}{r_{2}^{}},\,\frac{1}{r_{3}^{}}\Biggr)m_N^{} \;,
\hspace{5ex} r_{l}^{} \,\,=\,\, \frac{m_N}{M_l^{}} \;,
\end{eqnarray}
where, for convenience, we have introduced a scale parameter $m_N$ to represent the scale of the
heavy neutrino, which we choose to be the lightest of the heavy neutrinos.
The contribution to $\epsilon$ is given by
\begin{eqnarray}
\epsilon = U_{\nu N} U_{\nu N}^\dagger \approx U_0 U_0^\dagger \;.
\end{eqnarray}

We show three types of solutions relevant to our study of FCNC:
(a) sizeable $\epsilon_{12,13,23}$; (b) sizeable $\epsilon_{23}$ and small $\epsilon_{12,13}$;
and (c) sizeable $\epsilon_{13}$ and small $\epsilon_{12,23}$.
In case (a), the data from $\mu - e$ conversion in atomic nuclei constrain $|\epsilon_{12}|$
to be less than $1.7\times 10^{-7}$ which
makes $\epsilon_{13,23}$ too small to be of interest. We therefore need to find other classes of
solutions where $\epsilon_{12}$ is automatically much smaller than $\epsilon_{13,23}$. These are
the cases (b) and (c). If these types of solutions are correct, the constraints from $\tau$ decays
discussed previously in this paper are still relevant for experimental search.

The numerical results will be given by using the central values of
\,$\Delta m_{21}^2 = \bigl(7.65^{+0.23}_{-0.20}\bigr)\times
10^{-5}$\,eV$^2$\, and \,$\bigl|\Delta m_{31}^2\bigr|
=\bigl(2.40^{+0.12}_{-0.11}\bigr)\times 10^{-3}$\,eV$^2$,\,
determined by a recent fit to global neutrino
data~\cite{Schwetz:2008er}, and $U_{\rm PMNS}$ in the
tri-bimaximal form~\cite{Harrison:2002er} for simplicity
\begin{eqnarray} \label{Utb}
U_{\rm tribi}^{} \,\,= \left( \begin{array}{ccc} \displaystyle
\mbox{$\frac{-2}{\sqrt6}$} & \mbox{$\frac{1}{\sqrt3}$} & 0 \vspace{0.5ex} \\ \displaystyle
\mbox{$\frac{1}{\sqrt6}$} & \mbox{$\frac{1}{\sqrt3}$} & \mbox{$\frac{1}{\sqrt2}$}
\vspace{0.5ex} \\ \displaystyle
\mbox{$\frac{1}{\sqrt6}$} \,&\, \mbox{$\frac{1}{\sqrt3}$} \,&\, \mbox{$\frac{-1}{\sqrt2}$}
\end{array} \right) .
\end{eqnarray}
For the details of the following solutions, we refer to Ref.~\cite{ours}.

For case (a), a desired solution is given by
\begin{eqnarray}
&& U_0^a = U_{\rm PMNS} \left(
 \begin{array}{lll}
  a & a & i \sqrt{2} a \\
  b & b & i \sqrt{2} b \\
  c & c & i \sqrt{2} c
 \end{array} \right ) {\cal R} \;,\;\;\;
U_\delta^a = U_{\rm PMNS} \left(
 \begin{array}{lll}
  \delta_{11} & \delta_{12} & \delta_{13} \\
  \delta_{21} & \delta_{22} & \delta_{23} \\
  \delta_{31} & \delta_{32} & \delta_{33}
 \end{array}\right ) {\cal R} ~,
\end{eqnarray}
where ${\cal R} = {\rm diag} \left( \sqrt{r_1}, \sqrt{r_{2}}, \sqrt{r_{3}} \right)$.
There are two types of solutions corresponding to normal and inverted hierarchies in light
neutrino masses, but always one of the masses becomes zero as follows.
\\
(i) Normal hierarchy:
\begin{eqnarray}
&& a = 0 , \hspace{5ex}
\hat m_\nu = {\rm diag}\left(0,\, -1,\, \frac{c^2}{b^2}\right)2 \tilde\delta b m_N ~, \nonumber\\
&& \epsilon = \left(\begin{array}{ccc}
 0.33 &\, 0.33 - 0.97\,i \,&\, 0.33 + 0.97\,i \vspace{0.5ex} \\
 0.33 + 0.97\,i \,&  3.1   &   -2.5 +1.9\,i \vspace{0.5ex} \\
 0.33 - 0.97\,i   & -2.5-1.9 i  & 3.1
\end{array}\right) \bigl|b\bigr|^2\, r ~,
\end{eqnarray}
(ii) Inverted hierarchy:
\begin{eqnarray}
&& c = 0, \hspace{5ex}
\hat m_\nu = {\rm diag}\left(\frac{a^2}{b^2},\, -1,\, 0\right)2 \tilde\delta b m_N ~, \nonumber \\
&& \epsilon  = \left(\begin{array}{ccc}
 0.99 &\, 0.01-0.70\,i \,&\, 0.01-0.70\,i \vspace{0.5ex} \\
 0.01+0.70\,i \,& 0.50  & 0.50 \vspace{0.5ex} \\
 0.01+0.70\,i   & 0.50  & 0.50
\end{array}\right) |b|^2 ~ r ~,
\end{eqnarray}
where $\tilde \delta = \delta_{21} +\delta_{22} + i \sqrt{2}\delta_{23}$ and
$r = r_1 + r_{2} + 2 r_{3}$.
From $\mu - e$ conversion in atomic nuclei $\left(|\epsilon_{12}| = |\epsilon_{e \mu}|
< 1.7 \times 10^{-7} \right)$, $|b| \sqrt{r}$ is constrained
to be smaller than $4.1 \times 10^{-4}$ (normal hierarchy) or $4.9 \times 10^{-4}$ (inverted
hierarchy). In both cases, $|\epsilon_{13,23}|$ are constrained to be less than $O(10^{-7})$
which are way below the best constrained from $\tau \to \mu \bar \mu \mu$ and $\tau \to \pi^0 e$
decays.

\begin{figure}[t]
\centerline{\epsfig{figure=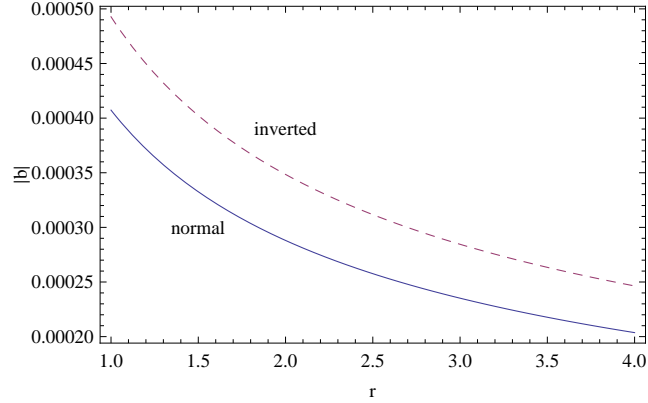, scale=1.0}}
\caption{For case (a), the upper limits on the magnitude of the element of $U_{\nu N}$
in terms of the heavy neutrino mass parameter $r \equiv r_1 + r_{2} + 2 r_{3}$.
The solid and dashed lines correspond to the normal and inverted hierarchy cases,
respectively.}
\label{fig1}
\end{figure}

In Fig.~\ref{fig1} we show the upper limits from the $\mu - e$ conversion constraint on the
magnitude of the element $b$ of $U_{\nu N}$ in terms of the heavy neutrino mass parameter
$r$.  Since $m_N$ is the lightest of $M_l$, $r$ is in the range $1 \leq r \leq 4$.
Depending on the heavy neutrino mass hierarchy, the value of $|b|$ can be different. With the
same constraint, to have the largest $b$, one would require the two heavier ones to be much
larger than the lightest $m_N$. As far as FCNC processes are concerned, the hierarchy of the
heavy neutrinos is not important because the parameter always involves $r$. But for the
production of a heavy lepton at LHC, via $q \bar q' \to W^* \to lN$ or
$q\bar q\to(Z^*,h^*)\to l E$ for example, it is preferred to have a larger $b$, because in
that case, not the combination $|b|^2 (r_1+r_{2}+2r_{3})$ but the individual $|b|^2 r_j$ is
relevant to the production cross section.

For case (b), the following form serves the purpose with the choice
$U_{\nu N}^{}=U_0^b + U_{\alpha\beta\gamma}^b + U_\delta^b$, where
\begin{eqnarray}
U_0^b \,\,=\, \left(\begin{array}{ccc} 0 & 0 & 0 \\
0 \,&\, a \,&\, ia \\ 0 & b & ib \end{array}\right) \!{\cal R}~, ~~~
U_{\alpha\beta\gamma}^b \,\,=\, \left(\begin{array}{ccc} \alpha \,&\, 0 & 0  \\
0 & 0 \,&\, 0  \\ 0 & 0 & 0 \end{array}\right) \!{\cal R}~, ~~~
U_\delta^b \,\,=\, \left(\begin{array}{ccc} 0 \,&\, \delta_{12}^{} \,&\, 0  \\
0 & \delta_{22}^{} & 0 \\ 0 & \delta_{32}^{} & 0 \end{array} \right) \!{\cal R} ~.
\end{eqnarray}
Here $\alpha$ is of order $[(a, b) \delta_{ij}]^{1/2}$ so that one should keep $\alpha^2$ terms
in the calculation, neglecting $\delta_{ij} \delta_{kl}$ and $\alpha \delta_{ij}$ terms.
The eigen-masses are
\begin{eqnarray}
\hat m_\nu \,\,=\,\, {\rm diag} \bigl( a\,\delta_{12}^{}-\alpha^2,\,
-2 a\,\delta_{12}^{}-\alpha^2,\, 0\bigr) m_N^{} ~,
\end{eqnarray}
and so this is an inverted hierarchy case with \,$m_{\nu_3}=0$.
Numerically, the matrix $\epsilon$ is given by
\begin{eqnarray}
\epsilon \,\,=\, \left(\begin{array}{ccc} 0 \,&\, 0 \,&\, 0 \vspace{0.5ex} \\
 0 & 1 & 1 \vspace{0.5ex} \\ 0 & 1 & 1
\end{array}\right) |a|^2\, \rho ~,
\end{eqnarray}
where $\rho = r_{2} + r_{3}$.
Thus, the constraint  \,$|\epsilon_{23}^{}|=|\epsilon_{\mu\tau}^{}|<4.9\times 10^{-4}$\,
from  \,$\tau\to\mu\bar\mu\mu$\, decays translates into
\,$|a|\sqrt{\rho} < 2.2\times 10^{-2}$. Since $r_1$ does not show up in $U^b_0$
in this case, it would be more convenient to choose $m_N$ to be the lightest of $M_{2,3}$.

For case (c), the desired results can be obtained by choosing
\,$U_{\nu N}^{}=U_0^c + U_{\alpha\beta\gamma}^c + U_\delta^b$,\,  with
\begin{eqnarray}
U_0^c \,\,=\, \left( \begin{array}{ccc} 0 \,&\, a \,&\, i a  \\
0 & 0 & 0  \\ 0 & b & i b \end{array}\right) \!{\cal R}~, ~~~
U_{\alpha\beta\gamma}^c \,\,=\, \left(\begin{array}{ccc} \alpha \,&\, 0 & 0  \\
\beta & 0 \,&\, 0  \\ 0 & 0 & 0 \end{array}\right) \!{\cal R}~.
\end{eqnarray}
This particular choice allows all the three light-neutrinos to have nonzero masses.
Taking $m_{\nu_2}=0.1$ eV, two possible solutions are found and give the matrix $\epsilon$
as follows.
\\
(i) Normal hierarchy $\bigl($with \,$m_{\nu_1}=0.0996$\,eV\, and \,$m_{\nu_3}=0.111$\,eV$\bigr)$:
\begin{eqnarray}
\epsilon \,\,=\,
\left(\begin{array}{ccc} 1 &\, 0 \,&\, 0.001-1.0\,i \vspace{0.5ex} \\
0 & 0 & 0 \vspace{0.5ex} \\ 0.001+1.0\,i \,& 0 & 1.1
\end{array}\right) |a|^2\, \rho ~,
\end{eqnarray}
(ii) Inverted hierarchy $\bigl($with \,$m_{\nu_1}=0.0996$\,eV\, and \,$m_{\nu_3}=0.0867$\,eV$\bigr)$:
\begin{eqnarray}
\epsilon \,\,=\,
\left(\begin{array}{ccc} 1 &\, 0 \,&\, 0.001+0.96\,i \vspace{0.5ex} \\
0 & 0 & 0 \vspace{0.5ex} \\ 0.001-0.96\,i \,& 0 & 0.93
\end{array}\right) |a|^2\, \rho ~.
\end{eqnarray}
The bound  \,$|\epsilon_{13}^{}|=|\epsilon_{e\tau}^{}|<4.2\times 10^{-4}$\, from
\,$\tau\to\pi^0 e$\, decays then implies  \,$|a|\sqrt{\rho} < 2.0\times10^{-2}$\,
in the two cases.

\begin{figure}[t]
\centerline{\epsfig{figure=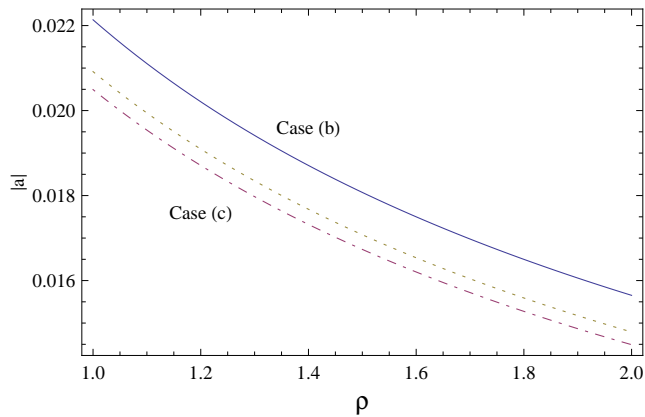, scale=1.0}}
\caption{For cases (b) and (c), the upper limits on the magnitude of the element of $U_{\nu N}$
in terms of the heavy neutrino mass parameter $\rho \equiv r_{2} + r_{3}$.
The solid line corresponds to case (b) and the dot-dashed and dotted lines correspond to the
normal and inverted hierarchy cases, respectively, in case (c).}
\label{fig2}
\end{figure}

In Fig.~\ref{fig2} we display the upper limits on the magnitude of the element $a$ of $U_{\nu N}$
in terms of the heavy neutrino mass parameter $\rho$ for cases (b) and (c).
In this case, $\rho$ is in the range $1 \leq \rho \leq 2$.
With the same constraint, the hierarchy that the heavier of $M_{2,3}$ is much larger than $m_N$
would be required to obtain the largest $a$.
Similarly to case (a), concerning FCNC processes, the hierarchy of the heavy neutrinos is
not important.  But, concerning the production of a heavy lepton $N$ or $E$ at LHC, a large cross
section can be obtained for $m_N \lesssim 115$ GeV~\cite{ours}.

The above examples clearly show that with the constraints from FCNC transitions as well as from the
tiny neutrino masses, the elements of $U_{\nu N}$ can still be large.  There is another class of
processes which also provides constraints on the elements of $U_{\nu N}$.
These processes involve neutral currents conserving lepton flavor and can be used to test
deviations from the SM predictions for electroweak precision data (EWPD)~\cite{ewpd}.
They have been measured mainly at LEP and provide bounds on the combinations of the diagonal
elements of $U_{\nu N}$. The constraints extracted from the EWPD are
$|(U_{\nu N})_{ii}|\le{\cal O}(0.01)$\,~\cite{ewpd}.
In contrast, the FCNC constraints discussed above involve combinations containing the off-diagonal
elements and impose more stringent constraints, such as
$|\epsilon_{12}| = \bigl|\sum_{k}(U_{\nu N})_{1k} (U_{\nu N}^*)_{2k} \bigr|<1.7\times 10^{-7}$.
The non-zero elements of $U_{\nu N}$ in the two examples we give above with
suppressed $\epsilon_{12}$, being at most of ${\cal{O}}(0.01)$, satisfy all these constraints.

Large elements of $U_{\nu N}$ also have important implications for a direct test of the model by
producing the heavy neutrinos at LHC.
The elements of $U_{\nu N}$ with the magnitude of order 0.01 are large enough to be detectable
at LHC~\cite{ours}.  The heavy neutrino $N$ can be produced through the mixing via
$q\bar q'\to W^*\to l^\pm N$.
Similarly, the heavy charged lepton $E$ can also be produced through the mixing via
$q\bar q\to(Z^*,h^*)\to l^\pm E^\mp$ and $q\bar q'\to W^*\to\nu E^\pm$.
At LHC the production cross section for a single heavy neutrino $N$ can be larger than 1 fb if
the heavy neutrino mass is less than 115 GeV with the elements of $U_{\nu N}$ being 0.01.
The production cross section of a single $E$ is slightly smaller. This can provide useful
information about this model.
\\


\noindent
{\bf Conclusions}

We have systematically studied various FCNC processes in the lepton sector in the framework of
Type III seesaw model.
Using the current experimental results, we have put the constraints on the parameters
$\epsilon_{l l'}$ which are responsible for tree level FCNC in the charged lepton sector.
The new processes that have been considered are: the LFV processes
$\tau \to P l$, $\tau \to V l$, $V \to l \bar l'$, $P \to l \bar l'$, $M \to M' l \bar l'$ and
muonium-antimuonium oscillation.

Although exchange both $Z$ and Higgs bosons at tree level can induce FCNC in charged lepton sector,
we find that there is no contribution from Higgs exchange in the processes $\tau \to P(V) l$ and
$V(P) \to l \bar l'$, and the effects of Higgs exchange are negligibly small in the last two classes
of processes.

We now compare constraints on various FCNC parameters obtained from processes considered in this work
with those obtained in previous studies.
It turns out that with currently available experimental data, the LFV processes considered in this work
involving $\tau$ leptons provide very stringent constraints on the FCNC parameter $\epsilon_{i \tau}$.
Our results show that the most stringent constraint on $\epsilon_{e\tau}$ comes from $\tau \to \pi^0 e$
decay with $|\epsilon_{e\tau}| < 4.2 \times 10^{-4}$.
$\tau \to \rho^0 \mu$ and $\tau \to \pi^0 \mu$ give very stringent constraints on $\epsilon_{\mu\tau}$
with $|\epsilon_{\mu\tau}| < 6.8\times 10^{-4}$ and $|\epsilon_{\mu\tau}| < 7.0\times 10^{-4}$,
respectively, comparable with $|\epsilon_{\mu\tau}| < 4.9\times 10^{-4}$ obtained from
$\tau \to \mu \bar \mu \mu$ in previous studies. The strongest constraint on $\epsilon_{e\mu}$ comes
from $\mu - e$ conversion in atomic nuclei studied previously with
$|\epsilon_{e\mu}| < 1.7 \times 10^{-7}$. The new constraint on $\epsilon_{e\mu}$ obtained from
processes considered in this work is much weaker.

Two body meson decays, such as $\Upsilon(3S) \to l \bar l'$,
$J/\Psi \to l \bar l'$, $\pi \to l \bar l'$ and $\eta^{(\prime)} \to l \bar l'$, provide
rather weak bounds on $|\epsilon_{l l'}|$ at most of order $10^{-1}$.
The constraints from semileptonic three body $B$ or $K$ decays of
the type $M \to M' l \bar l'$ are also rather weak with upper bounds on $|\epsilon_{l l'}|$ in the
range $O(10^{-2}) \sim O(1)$.

In the canonical seesaw models, where the elements of $U_{\nu N}$ are of the same order of magnitude
as that for an one generation seesaw model, $(m_\nu/m_N)^{1/2}$, it is not possible to have elements
of $\epsilon$ which are sufficiently large to reach the FCNC bounds studied in this paper. The FCNC
effects studied are therefore not interesting for the canonical seesaw models.
However, with more than one generation of light and heavy neutrinos, in certain special circumstances
the mixing is not constrained directly by the tiny neutrino masses and therefore can be large. Thus
in this class of seesaw models, it is possible to have large FCNC interactions.
These circumstances have been studied by several groups~\cite{large_mixing,ours}. We find some example
solutions which can lead to the FCNC parameters $\epsilon_{ij}$ large enough to reach the constraints
obtained here. The search for FCNC effects can still provide further information on the seesaw
models. We comment that efforts in constructing models with certain symmetries to evade the canonical
seesaw constraints on the mixing matrix $U_{\nu N}$ have been made in various ways~\cite{large_mix_model}.
It would be interesting to further study related phenomenology to test these models.

We would like to comment that in some processes considered in this work it is possible to have CP
violating signatures, such as lepton and  anti-lepton decay rate asymmetries, and asymmetries in
Z decays into $\bar l l'$ and $\bar l' l$~\cite{valle2}. To have non-zero effects, one needs not only
a weak phase appearing in CP violating couplings coming from the complex $\epsilon_{ij}$ and $U_{PMNS}$
matrix, but also a strong phase appearing in an absorptive part from loop induced decay amplitudes.
Since we consider that the heavy seesaw scale $M$ is heavier than $Z$, no absorptive part will be
developed with the heavy triplets in the loop. Only light degrees of freedom in the loop for
Z decays into $\bar l l'$ and $\bar l' l$ can generate the absorptive parts which are generally
small. The resulting CP violating effect will therefore be small. If polarizations of the initial
and final particles can be measured, it is possible to construct CP violating observables which
does not need the absorptive parts~\cite{hema}. We will carry out the detailed studies elsewhere.

Finally let us comment on several possible improvements on $\epsilon_{ll'}$ from future experiments.
Improved data for $\tau \to P l$ and $\tau \to V l$ decays at various facilities, such as B and $\tau$-Charm
factories, can improve
the bounds on $\epsilon_{ll'}$. Bounds from $V \to l\bar l'$ and $P \to l \bar l'$ can also be improved,
but may not be able to compete with constraints from other experiments.
The current bound from $B \to K \mu \tau$ is rather weak. But at LHCb about $10^{12}$ $b \bar b$ pairs are
expected to be produced each year, and this decay mode may be useful in improving bound on $\epsilon_{\mu\tau}$.
Rare kaon decays will be studied at J-PARC with high precisions so that the current weak bounds from kaon
decays may also become much stronger. But bounds obtained may still not be competitive with others.
$\mu- e$ conversion in atomic nuclei will also be studied at J-PARC with several orders of
magnitude improvement in sensitivity. Constraint on $\epsilon_{e\mu}$
can be improved by more than an order of magnitude. It may be very difficult to improve constraint on
$\epsilon_{e\mu}$ from muonium-antimuonium oscillation to the level $\mu - e$ conversion can achieve.
At present $Z \to l \bar l'$ do not provide the best bounds on $\epsilon_{ll'}$. However, the Giga-Z
modes at future colliders, such as ILC, the sensitivity can be improved by up to three orders of
magnitudes~\cite{ILC}. Future studies of $Z \to e \tau$ and $Z \to \mu \tau$ may improve the bounds
on $\epsilon_{e\tau}$ and $\epsilon_{\mu\tau}$.
It is clear that FCNC effects in Type III seesaw model can be further tested.
\\

\noindent
{\bf Acknowledgments}

This work was supported in part by the NSC and NCTS.


\end{document}